\documentclass[lettersize,journal]{IEEEtran}
\usepackage{amsmath,amsfonts,amssymb}
\usepackage{algorithmic}
\usepackage{algorithm}
\usepackage{array}
\usepackage[caption=false,font=normalsize,labelfont=sf,textfont=sf]{subfig}
\usepackage{textcomp}
\usepackage{stfloats}
\usepackage{url}
\usepackage{verbatim}
\usepackage{graphicx}
\usepackage{epstopdf}
\usepackage{cite}
\begin{document}

\title{Dynamic Event$-$Triggered Discrete$-$Time Linear Time$-$Varying System with Privacy$-$Preservation}

\author{Xuefeng Yang, Li Liu,~\IEEEmembership{Member,~IEEE,} Wenju Zhou,~\IEEEmembership{Member,~IEEE,} Jing Shi, Yinggang Zhang, Xin Hu and Huiyu Zhou}

\maketitle

\begin{abstract}
This paper focuses on discrete-time wireless sensor networks with privacy-preservation. In practical applications, information exchange between sensors is subject to attacks. For the information leakage caused by the attack during the information transmission process, privacy-preservation is introduced for system states. To make communication resources more effectively utilized, a dynamic event-triggered set-membership estimator is designed. Moreover, the privacy of the system is analyzed to ensure the security of the real data. As a result, the set-membership estimator with differential privacy is analyzed using recursive convex optimization. Then the steady-state performance of the system is studied. Finally, one example is presented to demonstrate the feasibility of the proposed distributed filter containing privacy-preserving analysis.
\end{abstract}


\section{Introduction}
\IEEEPARstart{D}{istributed} computing is the sharing of information among multiple pieces of software, which can run on a single machine or connecting by multiple computers over a network. Distributed computing applications are decomposed into multiple small parts and distributed to multiple computers for processing, which allows for reducing running time and sharing resource. Therefore, state estimation based on distributed computing has become a popular research topic \cite{ref1, ref2, ref3}.

Wireless sensor networks (WSNs) are mainly multi-hop self-organized distributed sensing networks, which are formed by large amount sensor nodes on the basis of wireless communication technology. Due to the advantage of the unrestricted formation, the uncertain network structure and the decentralized control among sensor nodes, WSNs are widely used in military \cite{ref4}, industrial \cite{ref5} and commercial  fields \cite{ref6}. However, performing efficient distributed processing is an extremely challenging topic, a great deal of studying on this topic are conducted, such as Kalman filter \cite{ref7,ref8} and $H\infty$ filter \cite{ref9,ref10}. Kalman filter is mainly applied to systems with deterministic noise or models \cite{ref11,ref12}. Considering the different applications in practice, to improve the performance of the traditional Kalman filter, novel methods are proposed, such as the unscented Kalman filter, the extended Kalman filter and the cubature Kalman filter \cite{ref13,ref14,ref15}. When the noise or system model is uncertain, $H\infty$ filter is applied to obtaining more accurate estimates and ensures the robustness of the system \cite{ref16,ref17,ref18,ref19}. However, in practical engineering applications, due to the noise complexity and modeling imprecision, the inaccurate values or even very terrible results are caused during the estimation process. The set-membership estimator (SME) is more suitable for solving such problems under unknown but bounded (UBB) noise \cite{ref20}. In recent years, the ellipsoid algorithm in the SME algorithm has been extensively studied \cite{ref21,ref22,ref23}. The ellipsoid algorithm mainly restricts the bounded noise to a set of ellipsoids. However, The ellipsoid algorithm is first transformed into a recursive convex optimization problem, which is then solved using interior point polynomials.

Information exchange between sensors is required in most estimation algorithms, which means that sensors need to broadcast information to their neighbors within a specific sampling period.  However, continuous or periodic information exchange between sensors consumes a lot of communication resources, which can even lead to network congestion or packet loss \cite{ref24,ref25}. Therefore, designing a feasible algorithm to decrease the frequency of data transmission between sensors is important for the sustainable use of communication resources. In practical research, event-triggered schemes (ETS) are divided into static event-triggered schemes (SETS) \cite{ref26} and dynamic event-triggered schemes (DETS) \cite{ref27,ref28,ref29}.
The threshold parameter of SETS is a fixed scalar, while DETS introduces an auxiliary parameter for the threshold. Auxiliary dynamic variables and dynamic threshold parameters are two typical algorithms for DETS. Since the DETS has higher resource utilization than SETS, DETS is more widely used.

Meanwhile, the advent of information era has infringed the network information security. During the information exchange, information tampering and leakage, the transmission procedure is confronted with the main threats to network communication \cite{ref30,ref31,ref32,ref33}. Similarly, the openness of interactive channels inevitably brings threats to information security. Therefore, it is necessary to protect the privacy of information. There are two main privacy-preserving methods: specially designed random noise and differential privacy methods. However, differential privacy methods have been widely studied due to the rigorous formulation derivation in the application and the proven security \cite{ref34,ref35}. Differential privacy methods protect the privacy of information using random states. It is very difficult for an attacker to deduce the real information. To prevent the data from being tampered and leaked during the exchange of sensor information, the filter estimation that incorporates privacy-preserving still needs to be further explored.

However, in reality, the existing models are unable to address several challenges at the same time. Firstly, the complexity of the noise leads to inaccurate models, despite simple assumptions for noise. It becomes a challenge to build more accurate models. Secondly, data transmission when sensors exchange information consumes too many resources, but it is difficulty to reduce the frequency of data transmission. Thirdly, the open nature of the information exchange channel poses a threat to data security. It is a challenge to ensure data security.

In summary, this paper studies a dynamic event-triggered discrete-time linear time-varying system (DTLVS) containing privacy-preservation. Considering the accuracy of the calculation and alleviating the complexity of the system, this paper uses an optimal bounding ellipsoid algorithm. The primary work is summarized as follows.

First, privacy-preserving noise is used to protect the initial state information from being stolen and leaked, and then analyze the privacy of the system. Secondly, the DETS is investigated to decrease the consumption of communication resources, and to achieve sustainable utilization of exchange resources. Finally, the event-triggered SME is designed to obtain accurate estimates even under the influence of uncertain noise. Note that the conditions required to be satisfied by the SME after introducing differential privacy are analyzed, so that one-step prediction state is always contained within the estimation ellipsoid.

The rest of the paper is summarized as the following parts. Section II establishes the DTLVS model under WSN and describes the design of an event-triggered SME. In section III, the conditions of the SME are presented after differential privacy. Based on this, the system stability is analyzed to verify its performance. Section IV simulates the proposed model, which can be efficiently applied to ship navigation. Section V summarizes the entire work.

Notation: $\mathbb{R}^{n}$ represents the  $n$-dimensional vector space, $\|.\|$ is the Euclidean norm, $\operatorname{col}_{N}\{.\} $ denotes the column vector consisting of $N$  blocks, $\operatorname{diag}_{N}\{.\}$ signifies the diagonal matrix consisting of $N$ blocks, $A 1 g(.)$ expresses the algorithm execution, and $\Xi $ indicates the execution domain of the algorithm.

\section{Main Work}
In a distributed WSN, the network topology is used for the model, which is defined as follows.

Let the index set of nodes in the directed graph be denoted as  $V=\{1,2, \ldots, N\} $ and an edge set of nodes be denoted as $\varepsilon \in V \times V$ . The topological graph is a weighted directed graph, $A=\left[a_{i j}\right] \in \mathbb{R}^{N \times N}$  expresses the adjacency matrix, where each element $a_{i j}$ represents the weight of edge between adjacent nodes. When data transfer is possible between two neighboring nodes, $a_{i j}>0$, otherwise, $a_{i j}=0$. The set of all neighbors for the node $i$ is denoted by $N_{i}=\{j \in V:(i, j) \in \varepsilon\} $. The directed graph consisting of $N$ nodes is denoted by $D=(V, \varepsilon, A)$.

Considering the noise from the external environment is described as UBB, and the ellipsoidal ensemble form is introduced as $\mathrm{Z} \triangleq\{a: a=b+E c,\|c\| \leq 1\} $, where the center of the ellipsoid is represented by $b \in \mathbb{R}^{N}$. Meanwhile, $E \in \mathbb{R}^{n \times m}$ is a lower triangular matrix satisfying $r a n k(E)=m \leq n$. Assuming that each element in the diagonal of matrix $E$ is greater than zero, the ellipsoidal representation can be rewritten as $\mathrm{Z} \triangleq\left\{a:(a-b)^{T} P^{-1}(a-b) \leq 1\right\} $, where $P=E E^{T}$ according to the cholesky decomposition.
\begin{figure}[!t]
\centering
{%
\resizebox*{8cm}{6cm}{\includegraphics{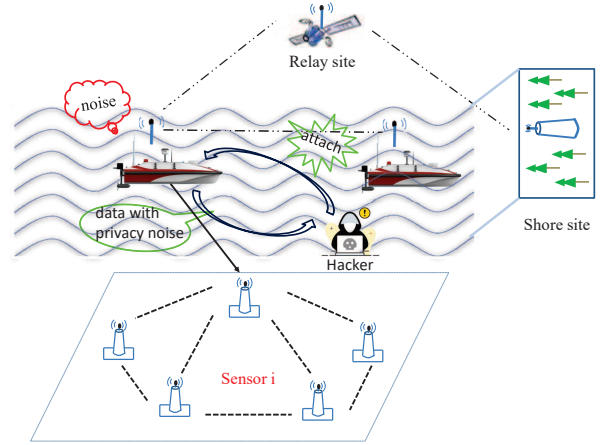}}}\hspace{5pt}
\caption{Transmission procedure of ship navigation system.} \label{fig1}
\end{figure}

\subsection{System Model}

In the field of ship navigation, some noise is inevitably generated because of unknown environmental changes during sailing. At the same time, the resistance generates changes so that the speed of the sailing ship is affected. Consequently, the data of the sailing speed and resistance, which are sensed or transmitted by the sensor, are containing noise. The disturbed data is processed by the sensor to obtain the original data. And as the sensors send and receive signals in a distributed manner, the open nature of their transmission channels may allow attacks to be made in the process causing distortion of the data. The security of the data is not guaranteed. Erroneous data is transmitted and sensed by neighboring sensors resulting in inaccurate data being obtained by the final estimator. It is necessary to ensure the security of its initial state. Consequently, introducing privacy noise protects the data security with mixing noise for the state value of the system state. The aim is to prevent data from being altered in the event of an attack. When the data is leaked, the data obtained by the stealer is the data containing privacy noise, nevertheless, the real original data cannot be obtained, i.e. the security of the original data is protected. The data transmission procedure of a ship navigation system is shown in Fig.~\ref{fig1}.

The system model of the ship navigation is established as follows.
\begin{equation}
\left\{\begin{array}{l}
\zeta_{k}=x_{k}+\eta_{k} \\
x_{k+1}=C_{k} \zeta_{k}+F_{k} w_{k}
\end{array}\right.,
\label{eq1}
\end{equation}
where $x_{k} \in \mathbb{R}^{\mathrm{n}_{x}}$ describes the state value. Affected by the
wind and waves, the ship sways. The ship swaying changes periodically due to the wave activities. $C_{k}$ and $F_{k}$ are defined as the time-varying periodic matrices, and $\zeta_{k} \in \mathbb{R}^{n_{\zeta}} $ describes the state after privacy-preserving, $\eta_{k} \in \mathbb{R}^{\mathrm{n}_{\eta}} $ describes the random privacy noise obeying the Laplace distribution. When the state information is leaked or stolen, the stealer obtains the information contains privacy noise, and the real state information cannot be obtained. This noise is represented as
\begin{equation}
\left\{\begin{array}{l}
\eta_{k} \sim \operatorname{Lap}(b) \\
b=c q^{k}
\end{array}\right.,
\label{eq2}
\end{equation}
where $c$ and $q$ satisfy the following conditions
\begin{equation}
\left\{\begin{array}{l}
c>0, \\
q \in(0,1).
\end{array}\right.
\label{eq3}
\end{equation}
$w_{k} \in \mathbb{R}^{\mathrm{n}_{w}}$ in Eq.(\ref{eq1}) expresses the process noise limited within a certain ellipsoid
\begin{equation}
W_{k} \triangleq\left\{w_{k}: w_{k}^{T}\left(R_{k}\right)^{-1} w_{k} \leq 1\right\},
\label{eq4}
\end{equation}
where $R_{k}=\left(R_{k}\right)^{T}>0$ expresses the real-valued matrix.

\subsection{Measurement Output Model}

The output measurement value of sensor $i$ at moment $k$ is expressed as follows
\begin{equation}
y_{k}^{i}=H_{k}^{i} x_{k}+D_{k}^{i} v_{k}^{i},
\label{eq5}
\end{equation}
where $y_{k}^{i} \in \mathbb{R}^{\mathrm{n}_{y}} $ expresses the output measurement of sensor $i$, $H_{k}^{i}$ and $D_{k}^{i}$ represent the time-varying coefficient matrices, and $v_{k}^{i} \in \mathbb{R}^{n_{v_{i}}}$ is measurement noise limited within a certain ellipsoid
\begin{equation}
V_{k}^{i} \triangleq\left\{v_{k}^{i}:\left(v_{k}^{i}\right)^{T}\left(Q_{k}^{i}\right)^{-1} v_{k}^{i} \leq 1\right\},
\label{eq6}
\end{equation}
where $Q_{k}^{i}=\left(Q_{k}^{i}\right)^{T}>0$ is the real-valued matrix.

Remark 1: For a known $\varsigma >0 $, when there exists
\begin{equation}
\left|x_{0}^{i_{0},2}-x_{0}^{i_{0},1}\right| \leq\left\{\begin{array}{l}
\varsigma , i=i_{0} \\
0, i \neq i_{0}
\end{array}\right.,
\label{eq7}
\end{equation}
the two sets of initial states $x_{0}^{i_{0},2}$, and $x_{0}^{i_{0},1}$ of system $i$ can be named $\varsigma >0$-adjacent \cite{ref36}.

Remark 2: Assuming that $\operatorname{Pr}\left[A \lg \left(x^{1}\right) \in \Xi\right] \leq e^{\varepsilon  \varsigma } \operatorname{Pr}\left[A \lg \left(x^{2}\right) \in \Xi\right] $ holds, adjacent initial states $x_{0}^{i,2}$ and $x_{0}^{i,1}$ can be obtained such that the algorithm meets $\epsilon$-differential privacy performance, i.e. $\epsilon_{i}=\frac{\varsigma  q_{i}}{c_{i}\left(q_{i}-\hat{A}_{i}\right)}$.

\subsection{Dynamic Event-Triggered Scheme}

To decrease the frequency of data transmission between sensors and improve the sustainable use of resources, a DETS \cite{ref37} is presented.

The event-triggered moment of the sensor $i$ is described as follows
\begin{equation}
t_{k+1}^{i}=\inf _{k \in \mathbb{N}}\left\{k>t_{k}^{i} \mid \theta_{i} l_{k}^{i}>\delta_{k}^{i}\right\},
\label{eq8}
\end{equation}
where
\begin{equation}
\left\{\begin{array}{l}
l_{i}=\left(h_{k}^{i}\right)^{T} \psi_{k}^{i} h_{k}^{i}-\sigma_{i}\left(\tilde{y}_{t_{k}^{i}}^{i}\right)^{T} \psi_{k}^{i} \tilde{y}_{t_{k}^{i}}^{i}, \\
h_{k}^{i}=\tilde{y}_{k}^{i}-\tilde{y}_{t_{k}^{i}}^{i}, \\
\tilde{y}_{k}^{i}=y_{k}^{i}-C_{k}^{i} \hat{x}_{k}^{i}.
\end{array}\right.
\label{eq9}
\end{equation}
Here $\sigma_{i} \in[0,1)$ denotes the specified threshold parameter, $\psi_{k}^{i}=\left(\psi_{k}^{i}\right)^{T}>0 $ denotes the sequence of weighting matrices to be determined, and $\tilde{y}_{k}^{i}$ denotes the measured residuals at moment $k$. $\delta_{k}^{i}$ is the key parameter and represents the auxiliary offset parameter, which satisfies Eq. (\ref{eq10})
\begin{equation}
\delta_{k}^{i}=\rho_{i} \delta_{k}^{i}-l_{k}^{i}.
\label{eq10}
\end{equation}
$\delta_{0}^{i} $ is the initial value of the auxiliary system. In Eqs. (\ref{eq8}) and (\ref{eq10}), $\rho_{i}$ and $\theta_{i}$ satisfy
\begin{equation}
\left\{\begin{array}{l}
0<\rho_{i}<1 \\
\theta_{i} \geq 1 / \rho_{i}
\end{array}\right..
\label{eq11}
\end{equation}

At the moment $k$, the packet scheduler $i$ is used to check the current packet $\left(k, \tilde{y}_{k}^{i}\right)$. When the event-triggered condition under Eq. (\ref{eq8}) is satisfied, the release moment $ t_{k+1}^{i}$ is calculated, at the same time, the packet at this moment is sent to the estimator $i$ and transmitted to its neighbors. Otherwise, this packet is discarded and the last reserved transmission packet $\left(t_{k}^{i}, \tilde{y}_{t_{k}^{i}}^{i}\right)$ is used. The auxiliary offset variable in Eq. (\ref{eq10}) is the key parameter in DETS, and then it can dynamically adjust the interval between two consecutive sampling moments, i.e. $(t_{k+1}^{i}-t_{k}^{i})$ .

\subsection{Event-Triggered Set-Membership Estimator}

State estimation needs  to achieve the full-scale confidence level. Therefore, this section aims to design an event-triggered set-membership estimator, and the one-step prediction for estimator $i$ is formulated as follows
\begin{equation}
\hat{x}_{k+1}^{i}=\hat{A}_{k}^{i} \hat{x}_{k}^{i}+\hat{B}_{k}^{i} \sum_{j \in N_{i}} a_{i j} \tilde{y}_{t_{k}^{j}}^{j},
\label{eq12}
\end{equation}
where $\tilde{k}_{j} \triangleq \arg \min _{\tilde{k}}\left\{k-t_{\tilde{k}}^{j} \mid k>t_{\tilde{k}}^{j}, \tilde{k} \in N\right\}$, $\hat{A}_{k}^{i}$ and $\hat{B}_{k}^{i}$ are the time-varying estimated gain matrices, while the initial state estimated value $\hat{x}_{0}^{i}$ satisfies
\begin{equation}
X_{0}^{i} \triangleq\left\{x_{0}:\left(x_{0}-\hat{x}_{0}^{i}\right)^{T}\left(U_{0}^{i}\right)^{-1}\left(x_{0}-\hat{x}_{0}^{i}\right) \leq \beta_{i}\right\},
\label{eq13}
\end{equation}
where $U_{0}^{i}=\left(U_{0}^{i}\right)^{T}>0$ and $\beta_{i}>0$ represent the time-varying real-valued matrix and the parameter variables of the ellipsoid, respectively.

In a system containing privacy noise $\eta_{k}$, UBB processes noise $w_{k}$ and measurement noise $v_{k} $, a confidence interval containing all estimates can be obtained by the set-membership estimator
\begin{equation}
X_{k+1}^{i} \triangleq\left\{x_{k+1}:\left(e_{k+1}^{i}\right)^{T}\left(U_{k+1}^{i}\right)^{-1} e_{k+1}^{i} \leq \beta_{i}\right\},
\label{eq14}
\end{equation}
where $e_{k+1}^{i}=x_{k+1}-\hat{x}_{k+1}^{i} $ denotes the estimation error and $U_{k+1}^{i}=\left(U_{k+1}^{i}\right)^{T}>0 $ is the time-varying real-valued matrix.

To satisfy the set-membership estimator with privacy-preserving, for a given sequence of scalars $\sigma_{i} \in[0,1)$, $\beta_{i}>0 $, $\rho_{i} $ and $\theta_{i} $ from Eq. (\ref{eq11}), set $\eta_{k} \in \mathbb{R}^{\mathrm{n}_{\eta}}$, $w_{k} \in \mathbb{R}^{\mathrm{n}_{w}}$ and $v_{k}^{i} \in \mathbb{R}^{\mathrm{n}_{v_{i}}} $, $i \in v $. Assuming that $U_{k+1}^{i}>0$, $\psi_{k}^{i}>0$, $\hat{A}_{k}^{i}$ and $\hat{B}_{k}^{i}$ exist, the one-step prediction state of $x_{k+1}$ can be resolved within the ellipsoid $X_{k+1}^{i}$ of the estimated state.

\section{Model Analysis}
The distributed system consists of $N$ subsystems, and the subsystems need to interact over communication. Therefore, to analyze the model, we restructure the parameters.
\begin{equation}
\begin{aligned}
\tilde{e}_{k}=\operatorname{col}_{N}\left\{e_{k}^{i}\right\}, \tilde{x}_{k}=\operatorname{col}_{N}\left\{x_{k}\right\}, \hat{x}_{k}=\operatorname{col}_{N}\left\{\hat{x}_{k}^{i}\right\}, \\
\tilde{\eta}_{k}=\operatorname{col}_{N}\left\{\eta_{k}\right\}, \tilde{M}_{k}=\operatorname{diag}_{N}\left\{M_{k}\right\}, M_{k}=2\left(b_{k}\right)^{2}, \\
\tilde{h}_{k}=\operatorname{col}_{N}\left\{h_{k}^{i}\right\}, \tilde{w}_{k}=\operatorname{col}_{N}\left\{w_{k}\right\}, \tilde{v}_{k}=\operatorname{col}_{N}\left\{v_{k}^{i}\right\}, \\
\tilde{\psi}_{k}=\operatorname{diag}_{N}\left\{\psi_{k}^{i}\right\}, \tilde{\Theta}=\operatorname{col}_{N}\left\{\theta_{i}\right\}, \tilde{\Sigma}=\operatorname{col}_{N}\left\{\sigma_{i}\right\}, \\
\tilde{\alpha}=\operatorname{col}_{N}\left\{\alpha_{i}\right\}, \tilde{\beta}=\operatorname{diag}_{N}\left\{\left(\beta_{i}\right)^{\frac{1}{2}}\right\}, \tilde{U}_{k}=\operatorname{diag}_{N}\left\{U_{k}^{i}\right\}, \\
\tilde{L}_{k}=\operatorname{diag}_{N}\left\{L_{k}^{i}\right\}, \tilde{R}_{k}=\operatorname{diag}_{N}\left\{R_{k}\right\}, \tilde{Q}_{k}=\operatorname{diag}_{N}\left\{Q_{k}^{i}\right\}, \\
\tilde{C}_{k}=\operatorname{diag}_{N}\left\{C_{k}\right\}, \tilde{F}_{k}=\operatorname{diag}_{N}\left\{F_{k}\right\}, \tilde{H}_{k}=\operatorname{diag}_{N}\left\{H_{k}^{i}\right\}, \\
\tilde{D}_{k}=\operatorname{diag}_{N}\left\{D_{k}^{i}\right\}, \hat{A}_{k}=\operatorname{diag}_{N}\left\{\hat{A}_{k}^{i}\right\}, \hat{B}_{k}=\operatorname{diag}_{N}\left\{\hat{B}_{k}^{i}\right\}.
\end{aligned}
\label{eq15}
\end{equation}

The following analysis focuses on the conditions guaranteeing steady-state performance, which satisfy the set-membership estimation after adding differential privacy.

\subsection{Analyzing Set-Membership Estimation with Differential Privacy}

The proposed set-membership estimator is designed to achieve DETS and privacy preservation, meanwhile it satisfies the one-step prediction state of $x_{k+1}$ being within estimation ellipsoid $X_{k+1}^{i} $.

Theorem 1: For a given scalar $\sigma_{i} \in[0,1) $, $\beta_{i}>0$, $\rho_{i}$ and $\theta_{i}$ satisfying Eq. \eqref{eq11}, set $\eta_{k} \in \mathbb{R}^{\mathrm{n}_{\eta}} $, $w_{k} \in \mathbb{R}^{\mathrm{n}_{w}}$ and $v_{k}^{i} \in \mathbb{R}^{\mathrm{n}_{v_{i}}} $, $i \in v$. Suppose that $U_{k+1}^{i}>0$, $\psi_{k}^{i}>0$, $\hat{A}_{k}^{i} $, $\hat{B}_{k}^{i} $ and a sequence of scalar $\epsilon_{k}^{m}>0$, $m=1,2,3,4$, there exists
\begin{equation}
\left(\begin{array}{cc}
-\tilde{U}_{k+1} & \Phi_{k} \\
* & \Lambda_{k}
\end{array}\right) \leq 0, \forall k \in \mathbb{N},
\label{eq16}
\end{equation}
where \[{{\Phi }_{k}}=\left[ ({{{\tilde{C}}}_{k}}-{{{\hat{A}}}_{k}}){{{\hat{x}}}_{k}},\tilde{\beta }({{{\tilde{C}}}_{k}}-{{{\hat{B}}}_{k}}A{{{\tilde{H}}}_{k}}){{L}_{k}},{{F}_{k}},{{{\tilde{C}}}_{k}},{{B}_{k}}A{{{\tilde{D}}}_{k}},{{{\hat{B}}}_{k}}A \right]\] and $\Lambda_{k}=\left[\Lambda_{k}^{p, q}\right]_{6 \times 6}$ are real-valued matrices. The non-zero terms contained in $\Lambda_{k}=\left[\Lambda_{k}^{p, q}\right]_{6 \times 6}$ are

\begin{equation}
\begin{aligned}
\Lambda_{k}^{1,1}=-\sum_{i=1}^{N} \beta_{i}+\epsilon_{k}^{1} N+\epsilon_{k}^{3} N+\epsilon_{k}^{4} N+\sum_{i=1}^{N} \delta_{k}^{i}, \\
\Lambda_{k}^{2,2}=-\epsilon_{k}^{4} I+\left(\tilde{L}_{k}\right)^{T}\left(\tilde{H}_{k}\right)^{T} \tilde{\beta}^{T} \tilde{\Theta} \tilde{\Sigma} \tilde{\Psi}_{k}\tilde{\beta} \tilde{H}_{k} \tilde{L}_{k}, \\
\Lambda_{k}^{2,5}=\left(\tilde{L}_{k}\right)^{T}\left(\tilde{H}_{k}\right)^{T} \tilde{\beta}^{T} \tilde{\Theta}\tilde{\Sigma} \tilde{\Psi}_{k} \tilde{D}_{k}, \\
\Lambda_{k}^{2,6}=-\left(\tilde{L}_{k}\right)^{T}\left(\tilde{H}_{k}\right)^{T} \tilde{\beta}^{T} \tilde{\Theta} \tilde{\Sigma}\tilde{\Psi}_{k}, \\
\Lambda_{k}^{3,3}=-\epsilon_{k}^{1}\left(\tilde{R}_{k}\right)^{-1},
\Lambda_{k}^{4,4}=\epsilon_{k}^{2}\left(\tilde{M}_{k}\right)^{-1}, \\
\Lambda_{k}^{5,5}=-\epsilon_{k}^{3}\left(\tilde{Q}_{k}\right)^{-1}+\left(\tilde{D}_{k}\right)^{T} \tilde{\Theta}\tilde{\Sigma} \tilde{\Psi}_{k} \tilde{D}_{k}, \\
\Lambda_{k}^{5,6}=-\left(\tilde{D}_{k}\right)^{T} \tilde{\Theta} \tilde{\Sigma}\tilde{\Psi}_{k},
\Lambda_{k}^{6,6}=\tilde{\Theta}\left(\tilde{\Sigma}-I\right) \tilde{\Psi}_{k}.
\end{aligned}
\label{eq17}
\end{equation}

Proof: First, to simplify Eq. \eqref{eq13}, $\left(e_{0}^{i}\right)^{T}\left(U_{0}^{i}\right)^{-1} e_{0}^{i} \leq \beta_{i}$ is obtained. If there exists $x_{k} \in X_{k}^{i} $ satisfying $\left(e_{k}^{i}\right)^{T}\left(U_{k}^{i}\right)^{-1} e_{k}^{i} \leq \beta_{i}$, it is sufficient to prove that $x_{k+1} \in X_{k+1}^{i}$ exists and $\left(e_{k+1}^{i}\right)^{T}\left(U_{k+1}^{i}\right)^{-1} e_{k+1}^{i} \leq \beta_{i} $ holds.

An alternative formulation of the ellipsoid, where the estimation error can be obtained from $e_{k+1}^{i}=x_{k+1}-\hat{x}_{k+1}^{i} $ at time $k+1$, expressed as $\left(x_{k}-\hat{x}_{k}^{i}\right)^{T}\left(U_{k}^{i}\right)^{-1}\left(x_{k}-\hat{x}_{k}^{i}\right) \leq \beta_{i}$.

To simplify the set of ellipsoids where the estimated errors are located $\beta_{i}-\left(e_{k}^{i}\right)^{T}\left(U_{k}^{i}\right)^{-1} e_{k}^{i} \geq 0$, which may be transmitted as $\beta_{i}^{-1} e_{k}^{i}\left(e_{k}^{i}\right)^{T} \leq U_{k}^{i}$ under the Schur complementary. Then, we decompose $U_{k}^{i}$ using the cholesky factorization, i.e. $U_{k}^{i}=L_{k}^{i}\left(L_{k}^{i}\right)^{T}$, where $L_{k}^{i} $ is a lower triangular matrix satisfying all elements on the diagonal are positive. Therefore, $\left(\beta_{i}\right)^{-1} e_{k}^{i}\left(e_{k}^{i}\right)^{T} \leq L_{k}^{i}\left(L_{k}^{i}\right)^{T}$ is redefined.
Let $\alpha_{i}=\left(\beta_{i}\right)^{-\frac{1}{2}}\left(L_{k}^{i}\right)^{-1}\left(x_{k}-\hat{x}_{k}^{i}\right) $, then
\begin{equation}
\alpha_{i}^{T} \alpha_{i}=\left(\beta_{i}\right)^{-1}\left(x_{k}-\hat{x}_{k}^{i}\right)^{T}\left(U_{k}^{i}\right)^{-1}\left(x_{k}-\hat{x}_{k}^{i}\right) \leq 1, \\
\label{eq18}
\end{equation}
i.e. meeting $\left\|\alpha_{i}\right\| \leq 1 $.

From Eq. \eqref{eq18}, the state is rewritten as
\begin{equation}
x_{k}=\left(\beta_{i}\right)^{\frac{1}{2}} L_{k}^{i} \alpha_{i}+\hat{x}_{k}^{i}.
\label{eq19}
\end{equation}

Subsequently, the one-step prediction state estimation error $e_{k+1}^{i}$ is rewritten from Eqs. \eqref{eq1}, \eqref{eq12} and \eqref{eq19}
\begin{equation}
\begin{aligned}
e_{k+1}^{i}=C_{k}\left(x_{k}+\eta_{k}\right)+F_{k} w_{k}-\left(\hat{A}_{k}^{i} x_{k}^{j}+\hat{B}_{k}^{i} \sum_{j \in N_{i}} a_{i j} \tilde{y}_{t_{\tilde{k}_{i}}^{j}}^{j}\right) \\
=\left(C_{k}-\hat{A}_{k}^{i}\right) \hat{x}_{k}^{j}+\left(\beta_{i}\right)^{\frac{1}{2}} C_{k} L_{k}^{i} \alpha_{i}+F_{k} w_{k}+C_{k} \eta_{k} \\
-\hat{B}_{k}^{i} \sum_{j \in N_{i}} a_{ij}\left(\beta_{j}\right)^{\frac{1}{2}} H_{k}^{j} L_{k}^{j} \alpha_{j}-\hat{B}_{k}^{i} \sum_{j \in N_{i}} a_{i j}\left(D_{k}^{j} v_{k}^{j}-h_{k}^{j}\right).
\end{aligned}
\label{eq20}
\end{equation}

Let $\psi_{k}=\left[1, \tilde{\alpha}, \tilde{w}_{k}, \tilde{\eta}_{k}, \tilde{v}_{k}, \tilde{h}_{k}\right]^{T} $, then Eq. \eqref{eq20} is represented as $\tilde{e}_{k+1}=\Phi_{k} \psi_{k}$, where $\Phi_{k}=\left[\left(\tilde{C}_{k}-\hat{A}_{k}\right) \hat{x}_{k}, \tilde{\beta}\left(\tilde{C}_{k}-\hat{B}_{k} A \tilde{H}_{k}\right) L_{k}, \tilde{F}_{k}, \tilde{C}_{k},-\hat{B}_{k} A \tilde{D}_{k}, \hat{B}_{k} A\right]$. Thus, the one-step prediction state error $\left(e_{k+1}^{i}\right)^{T}\left(U_{k+1}^{i}\right)^{-1} e_{k+1}^{i} \leq \beta_{i}$ can be rewritten as
\begin{equation}
\psi_{k}^{T}\left(\Phi_{k}^{T} U_{k+1}^{-1} \Phi_{k}+\Theta\right) \psi_{k} \leq 0,
\label{eq21}
\end{equation}
where $\Theta=\operatorname{diag}\left\{-\sum_{i=1}^{N} \beta_{i}, 0,0,0,0,0\right\} $.

From Eqs. \eqref{eq2}, \eqref{eq4}, \eqref{eq6} and $\left\|\alpha_{i}\right\| \leq 1$, such that $\psi_{k}^{T} \Gamma_{k}^{1} \psi_{k} \geq 0$, $\psi_{k}^{T} \Gamma_{k}^{2} \psi_{k} \geq 0$, $\psi_{k}^{T} \Gamma_{k}^{3} \psi_{k} \geq 0$ and $\psi_{k}^{T} \Gamma_{k}^{4} \psi_{k} \geq 0$, in which $\Gamma_{k}^{1}=\operatorname{diag}\left\{N, 0,-\tilde{R}_{k}^{-1}, 0,0,0\right\} $, $\Gamma_{k}^{2}=\operatorname{diag}\left\{0,0,0, \tilde{M}_{k}^{-1}, 0,0\right\} $, $\Gamma_{k}^{3}=\operatorname{diag}\left\{N, 0,0,0,-\tilde{Q}_{k}^{-1}, 0\right\}$, and $\Gamma_{k}^{4}=\operatorname{diag}\{N,-I, 0,0,0,0\} $.

According to Eqs. \eqref{eq8} and \eqref{eq9}, the inequality can be obtained
\begin{equation}
\delta_{k}^{i}-\theta_{i}\left(\left(h_{k}^{i}\right)^{T} \psi_{k}^{i} h_{k}^{i}-\sigma_{i}\left(\tilde{y}_{k}^{i}-h_{k}^{i}\right)^{T} \psi_{k}^{i}\left(y_{k}^{i}-h_{k}^{i}\right)\right) \geq 0.
\label{eq22}
\end{equation}

Note that,
\begin{equation}
\begin{aligned}
\delta_{k}^{i}-\theta_{i}\left(\left(h_{k}^{i}\right)^{T} \psi_{k}^{i} h_{k}^{i}-\sigma_{i}\left(\tilde{y}_{k}^{i}-h_{k}^{i}\right)^{T} \psi_{k}^{i}\left(y_{k}^{i}-h_{k}^{i}\right)\right)\\
=\delta_{k}^{i}+\theta_{i} \sigma_{i}\left(h_{k}^{i}\right)^{T} \psi_{k}^{i} h_{k}^{i}-\theta_{i}\left(h_{k}^{i}\right)^{T} \psi_{k}^{i} h_{k}^{i} \\
+\theta_{i} \sigma_{i} \alpha_{i}^{T}\left(L_{k}^{i}\right)^{T}\left(\beta_{i}^{1 / 2}\right)^{T}\left(H_{k}^{i}\right)^{T} \psi_{k}^{i} H_{k}^{i} \beta_{i}^{1 / 2} L_{k}^{i} \alpha_{i}\\
+\theta_{i} \sigma_{i} \alpha_{i}^{T}\left(L_{k}^{i}\right)^{T}\left(\beta_{i}^{1 / 2}\right)^{T}\left(H_{k}^{i}\right)^{T} \psi_{k}^{i} D_{k}^{i} v_{k}^{{i}} \\
+\theta_{i} \sigma_{i}\left(v_{k}^{i}\right)^{T}\left(D_{k}^{i}\right)^{T} \psi_{k}^{i} H_{k}^{i} \beta_{i}^{1 / 2} L_{k}^{i} \alpha_{i}\\
+\theta_{i} \sigma_{\dot{i}}\left(v_{k}^{i}\right)^{T}\left(D_{k}^{i}\right)^{T} \psi_{k}^{i} D_{k}^{i} v_{k}^{i}-\theta_{i} \sigma_{i}\left(h_{k}^{i}\right)^{T} \psi_{k}^{i} H_{k}^{i} \beta_{i}^{1 / 2} L_{k}^{i} \alpha_{i} \\
-\theta_{i} \sigma_{i}\left(h_{k}^{i}\right)^{T} \psi_{k}^{i} D_{k}^{i} v_{k}^{i}-\theta_{i}\sigma_{i}\left(v_{k}^{i}\right)^{T}\left(D_{k}^{i}\right)^{T} \psi_{k}^{i} h_{k}^{i}\\
-\theta_{i} \sigma_{i}\alpha_{i}^{T}\left(L_{k}^{i}\right)^{T}\left(\beta_{i}^{1 / 2}\right)^{T}\left(H_{k}^{i}\right)^{T} \psi_{k}^{i} h_{k}^{i} .
\end{aligned}
\label{eq23}
\end{equation}
The nonzero terms of matrix $\Xi_{k}=\left[\Xi_{k}^{p, q}\right]_{6 \times 6}$ from Eq. \eqref{eq23} are defined as $\Xi_{k}^{1,1}=\sum_{i=1}^{N} \delta_{k}^{i} $, $\Xi_{k}^{2,2}=\Lambda_{k}^{2,2}+\epsilon_{k}^{4} I $, $\Xi_{k}^{2,5}=\Lambda_{k}^{2,5}$, $\Xi_{k}^{2,6}=\Lambda_{k}^{2,6}$, $\Xi_{k}^{5,5}=\Lambda_{k}^{5,5}+\epsilon_{k}^{3} \tilde{Q}_{k}^{-1} $, $\Xi_{k}^{5,6}=\Lambda_{k}^{5,6} $ and $\Xi_{k}^{6,6}=\Lambda_{k}^{6,6} $.

According to the S-procedure, assuming that a sequence of scalars $\epsilon_{k}^{m}>0$, i.e. $m=1,2,3,4$ exists and all of them are positive, inequality Eq. \eqref{eq21} can be rewritten as
\begin{equation}
\Phi_{k}^{T} \tilde{U}_{k+1}^{-1} \Phi_{k}+\Theta+\epsilon_{k}^{1} \Gamma_{k}^{1}+\epsilon_{k}^{2} \Gamma_{k}^{2}+\epsilon_{k}^{3} \Gamma_{k}^{3}+\epsilon_{k}^{4} \Gamma_{k}^{4}+\Xi_{k} \leq 0.
\label{eq24}
\end{equation}
The inequality Eq. \eqref{eq24} is simplified using the Schur complement to obtain Eq. \eqref{eq16}.

\subsection{Analysis of System State Stability }

The definited $g_{k}=E\left[x_{k} x_{k}^{T}\right]$ satisfies the following equation
\begin{equation}
g_{k+1}=C_{0} g_{k} C_{0}^{T}+C_{0} M_{k} C_{0}^{T}+F_{0} R_{k} F_{0}^{T}.
\label{eq25}
\end{equation}

Theorem 2: For the system in Eqs. \eqref{eq1} and \eqref{eq5}, the solution $g_{k}$ with any initial value $g_{0} \geq 0$ converges to a unique semi-positive definite solution $g$ of the Lyapunov equation. Note that the matrix $C$ is stable.
\begin{equation}
g=C_{0} g C_{0}^{T}+C_{0} M_{0} C_{0}^{T}+C_{0} R_{0} C_{0}^{T}.
\label{eq26}
\end{equation}
In addition, $\lim _{k \rightarrow \infty} g_{k}=g$, $\lim _{k \rightarrow \infty} R_{k}=R$, $\lim _{k \rightarrow \infty} Q_{k}=Q$ and $\lim _{k \rightarrow \infty} M_{k}=M$ are held.

Proof: Since matrix $C$ is stable, the spectral radius $\rho (C)<1$ of matrix $C$ is obtained. In this context, $g=\lim _{k \rightarrow \infty} g_{k} $. In addition, $\lim _{k \rightarrow \infty} R_{k}=R$, $\lim _{k \rightarrow \infty} Q_{k}=Q $, $\lim _{k \rightarrow \infty} M_{k}=M$ are held.  So when $g_{k} $ converges to the unique semi-positive definite solution $g$ of the Lyapunov equation, the system has the steady-state performance.

\section{Numerical Simulation}
This section applies the model to a ship navigation system, which has the similar description in \cite{ref38}, where a DTLVS can be modeled from all external factors
\begin{equation}
{\left[\begin{array}{l}
x_{k+1}^{1} \\
x_{k+1}^{2}
\end{array}\right]=\left[\begin{array}{ll}
0.9653 & -0.0021 \\
-0.054 & 0.7654+0.2 * \sin (k)
\end{array}\right]\left[\begin{array}{l}
x_{k}^{1}+\eta_{k}^{1} \\
x_{k}^{2}+\eta_{k}^{2}
\end{array}\right]}, \\
\label{eq27}
\end{equation}
where $x_{k}^{1}$ and $x_{k}^{2}$ denote the speed and resistance in ship navigation, respectively. In the actual ship navigation, the noise is mainly from the disturbance of the external environment, such as unpredictable weather, waves, etc. Note that the privacy noise $\eta_{k}$ is random to protect the system safety. Therefore, the time-varying matrix is defined as
\begin{equation}
C_{k}=\left[\begin{array}{cc}
0.9653 & -0.0021 \\
-0.054 & 0.7654+0.2 * \sin (k)
\end{array}\right],
\label{eq28}
\end{equation}
\begin{equation}
F_{k}=\left[\begin{array}{c}
0.22+0.22* \sin (k) \\
0.22
\end{array}\right].
\label{eq29}
\end{equation}

\begin{figure}[h]
\centering
{%
\resizebox*{5cm}{!}{\includegraphics{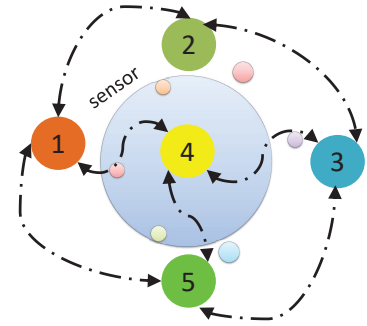}}}\hspace{5pt}
\caption{Sensor relationship diagram.} \label{fig2}
\end{figure}

To provide more accurate and convincing data in the simulation, a network topology with five sensor nodes is designed, as shown in Fig.~\ref{fig2}. Each sensor receives and transmits data only to its neighboring sensors. Sensor nodes $i$ and $j$ exchange information to receive or transmit data when the element $a_{i j}$ in the adjacency matrix is set as 1. The sensor is affected by the measurement noise $v_{k}^{i}$ while transmitting data, where the time-varying coefficient matrices are $H_{k}^{i}=\left[\begin{array}{ll}
0 & 1.1+0.11^{*}(i+1)-0.11^{*} \sin (k)
\end{array}\right]
$ and $D_{k}^{i}=1 /(i+1)$.	

In this section, let the initial speed and initial resistance of the ship sailing be 1.7m/s and 3.7kgf. Set the estimated values of the initial moment as $\hat{x}_{1}^{0}=\left[\begin{array}{ll}
1.8 & 3.7
\end{array}\right]
$, $\hat{x}_{2}^{0}=\left[\begin{array}{ll}
1.6 & 3.8
\end{array}\right]
$, $\hat{x}_{3}^{0}=\left[\begin{array}{ll}
1.8 & 3.55
\end{array}\right]$
, $\hat{x}_{4}^{0}=\left[\begin{array}{ll}
1.35 & 3.3
\end{array}\right]
$ and $\hat{x}_{5}^{0}=\left[\begin{array}{ll}
1.5 & 3.9
\end{array}\right]
$. Let $U_{0}=\operatorname{diag}_{2}\{40 \quad 40\}$, $\beta_{i}=1$, $R_{k}=0.4 $, $Q_{k}^{i}=0.2$ and $b=1$.
 \begin{figure*}[!t]
\centering
\subfloat[Ship sailing speed.]{%
\resizebox*{7cm}{!}{\includegraphics{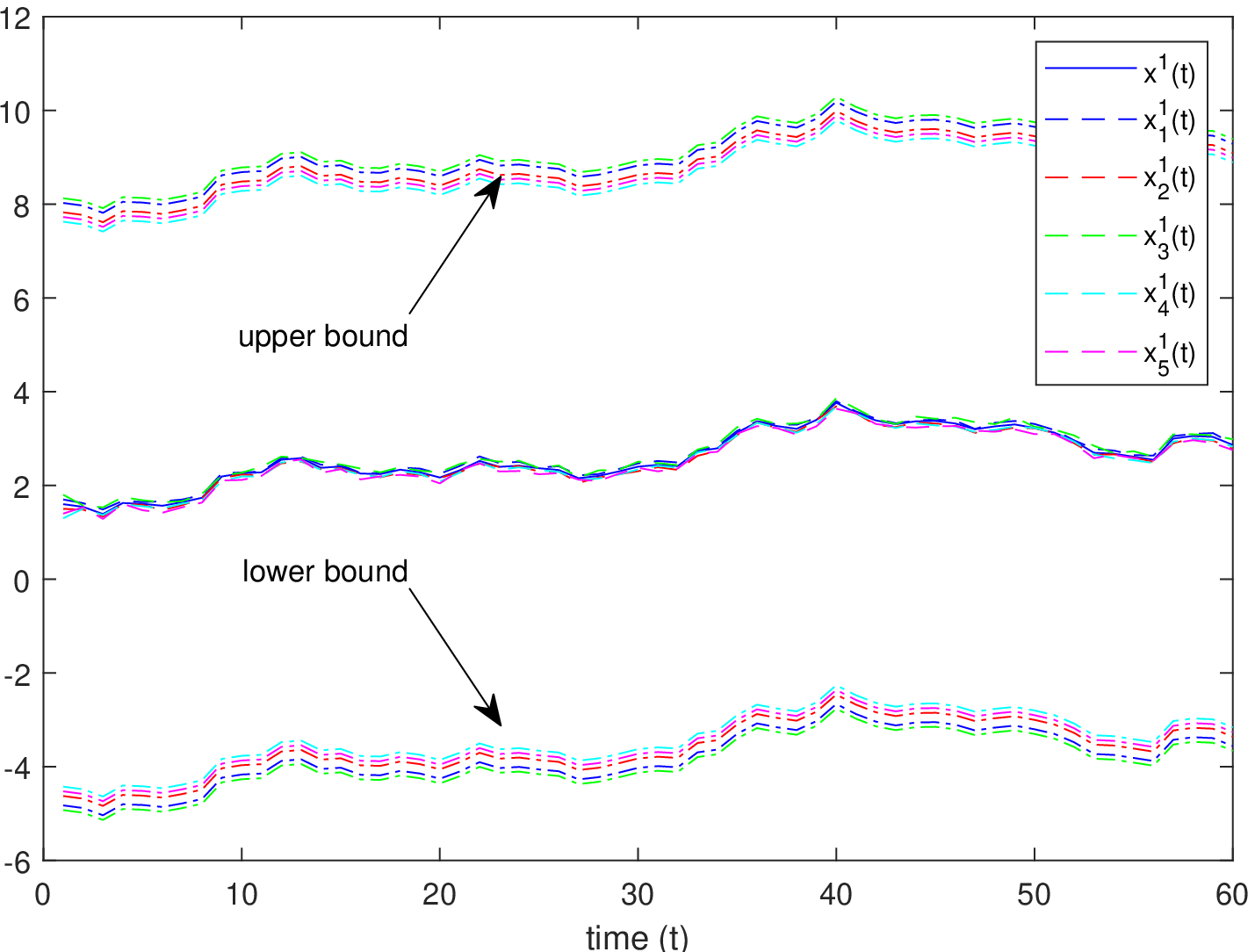}}}
\hfil
\subfloat[Partial value of ship sailing speed.]{%
\resizebox*{7cm}{!}{\includegraphics{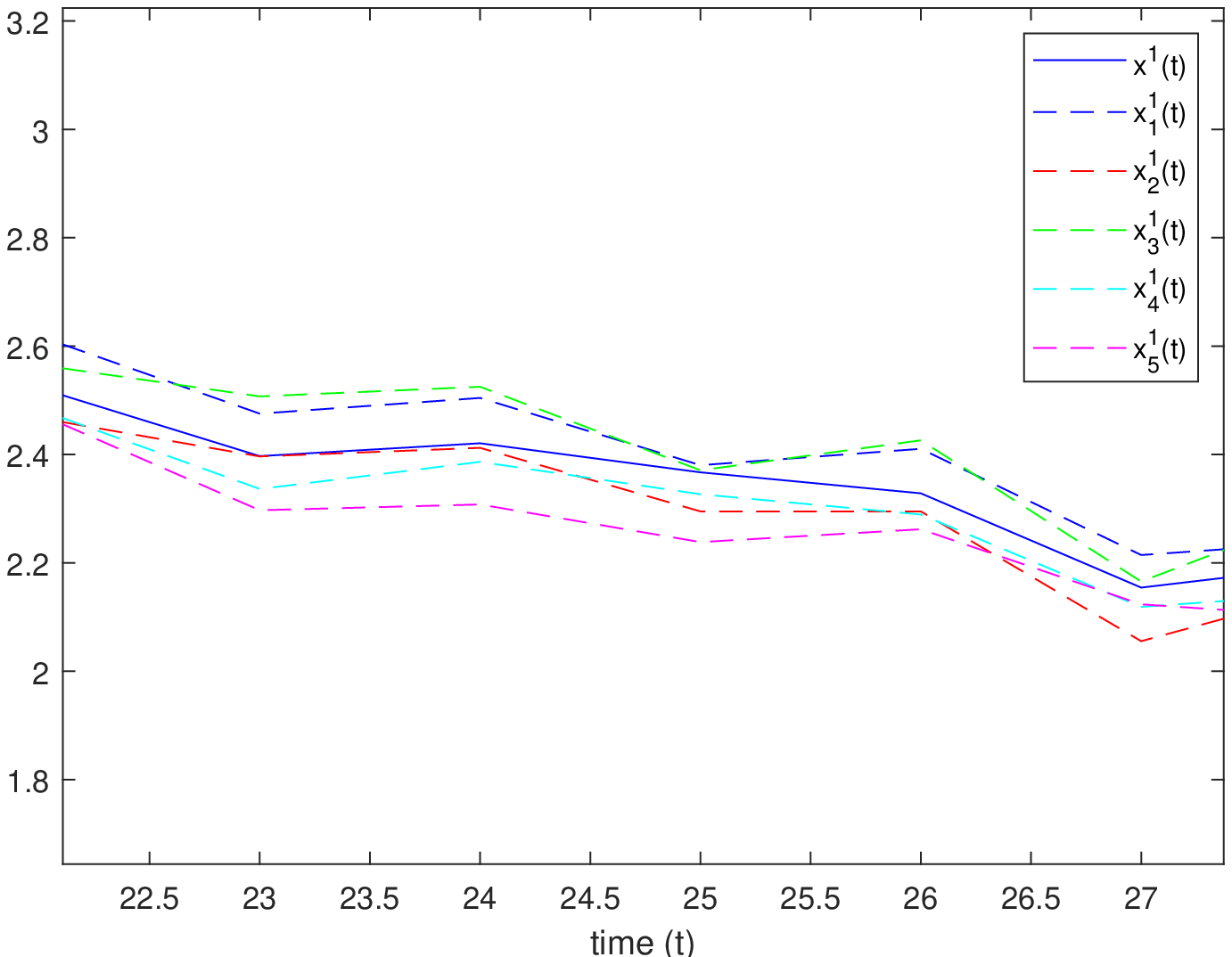}}}\hspace{5pt}
\subfloat[Ship sailing resistance.]{%
\resizebox*{7cm}{!}{\includegraphics{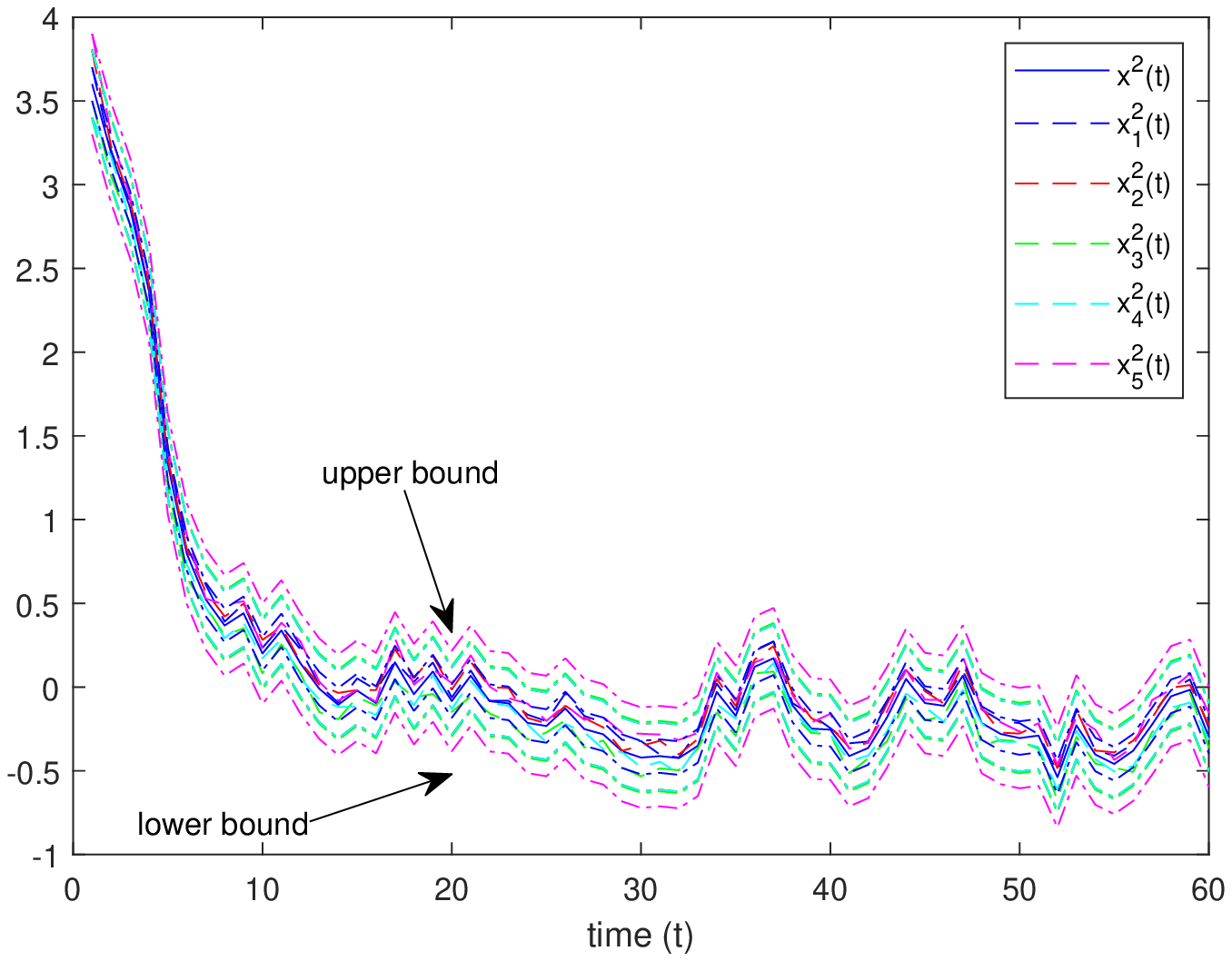}}}
\hfil
\subfloat[Partial value of ship sailing resistance.]{%
\resizebox*{7cm}{!}{\includegraphics{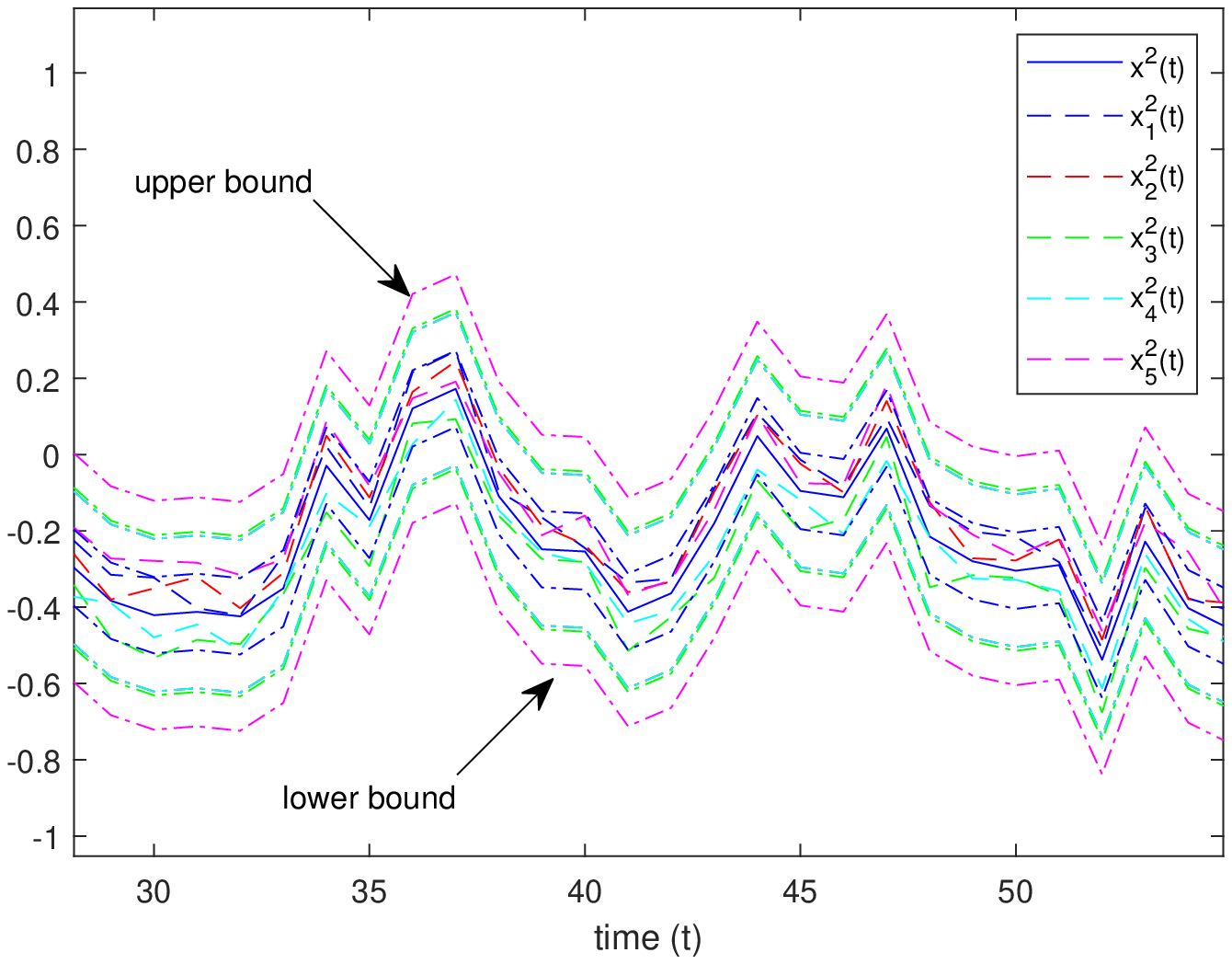}}}
\caption{System state estimation for speed and resistance.} \label{fig3}
\end{figure*}

\begin{figure*}[!t]
\centering
\subfloat[Ship sailing speed.]{%
\resizebox*{7cm}{!}{\includegraphics{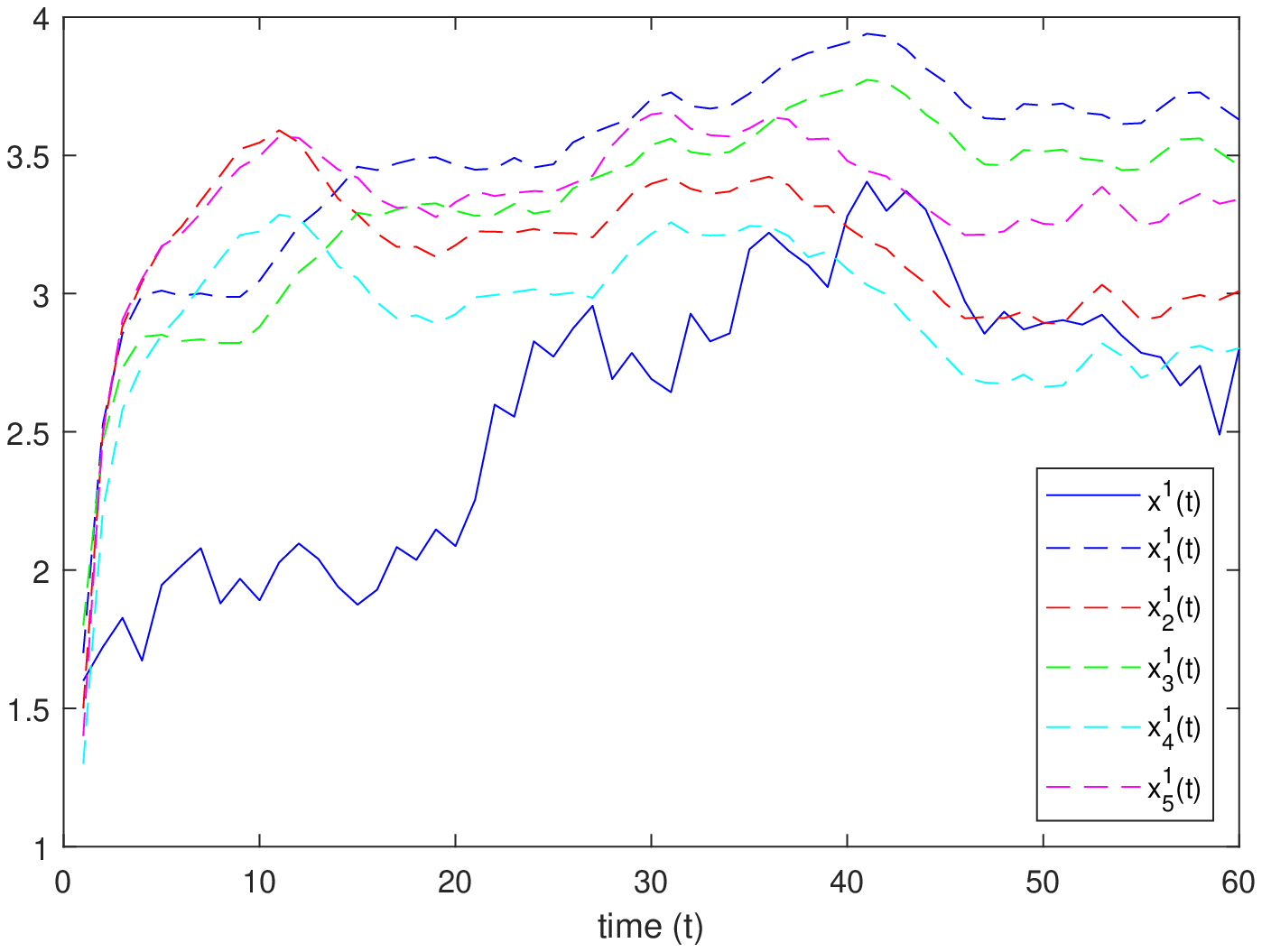}}}\hspace{5pt}
\hfil
\subfloat[Ship sailing resistance.]{%
\resizebox*{7cm}{!}{\includegraphics{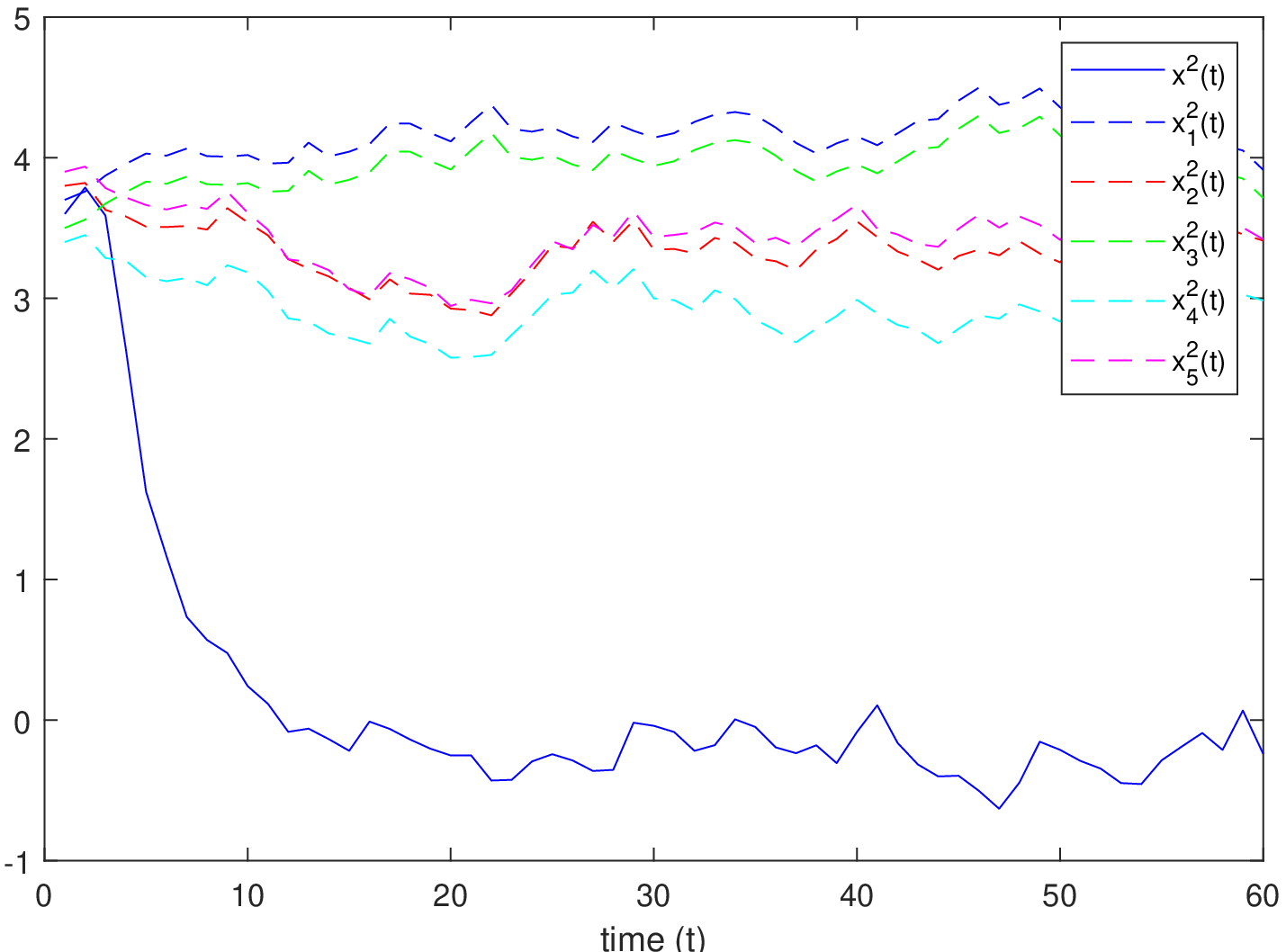}}}
\caption{Comparison for simulation of speed and resistance.} \label{fig4}
\end{figure*}

In the DETS model, set the values of its corresponding parameters as follows: $\sigma_{i}=\{0.98,0.9,0.8,0.85,0.93\}$, $\rho_{i}=0.7$, $\theta_{i}=30$ and $\delta_{i}^{0}=\{0.25,0.2,0.15,0.1,0 .05\} $, where $i=1,2,3,4,5$.

In Figs.~\ref{fig3} and~\ref{fig4}, $x_{k}^{1}$, $x_{k}^{2}$ denote the sailing speed and resistance parameters in the state variables, and $x_{k}^{i, 1}$, $x_{k}^{i, 2}$ denote the estimation value of the sailing speed and resistance. The designed set-membership estimation model with privacy-preserving is simulated and compared with the model in \cite{ref39}. The actual value is always in the estimation interval, and the estimated value obtained by the estimator is close to the actual value given in Fig.~\ref{fig3}. From Fig.~\ref{fig4}, when the other models are applied to the ship navigation system, there is a certain disparity between the estimated value and the actual value. And there is a certain difference in the numerical fluctuation of the speed and resistance of the ship. The actual state value of the system cannot be judged by the estimator. Therefore, the estimator designed in this paper can be well applied in the ship navigation system.
\begin{figure}
\centering
{%
\resizebox*{7cm}{!}{\includegraphics{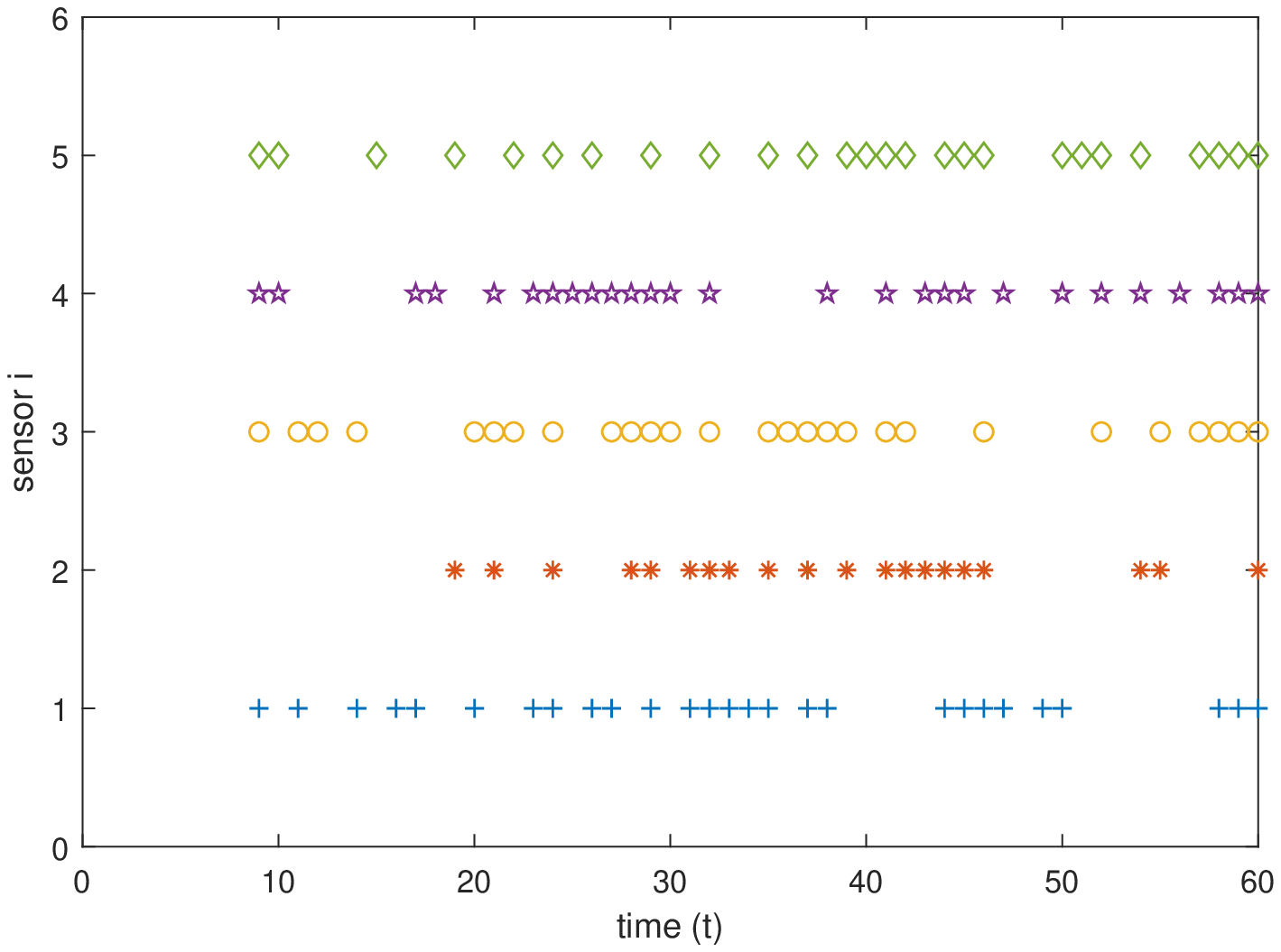}}}\hspace{5pt}
\caption{Dynamic event-triggered scheme for time interval.} \label{fig5}
\end{figure}
\begin{figure}
\centering
{%
\resizebox*{7cm}{!}{\includegraphics{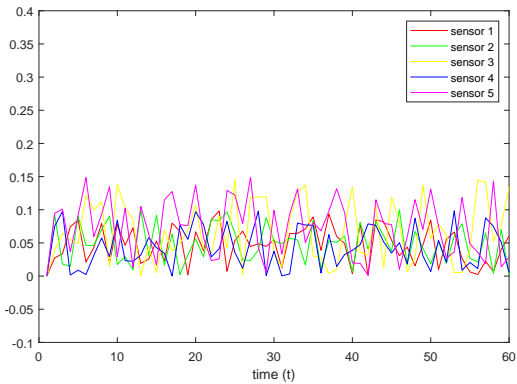}}}\hspace{5pt}
\caption{Estimation error of each sensor.} \label{fig6}
\end{figure}

Adding DETS to the ship application, the release time interval of specific events is shown in Fig.~\ref{fig5}. It shows that the dynamic event triggering scheme is applied, the triggering time is discontinuous, and the number of event-triggered events is significantly decrease. The DETS is able to better decrease the frequency of data transmission between sensors and their neighbors, and then, improve the sustainable use of resources.

From Fig.~\ref{fig6}, the estimation error values of each sensor can be fluctuated within a certain range, and the fluctuation range is very small. It may be obtained that the model designed can be efficiently applied in the ship navigation system, and optimal estimation values can be obtained at each moment.

\section{Conclusion}
This paper addressed information privacy-preservation for DTLVS. The initial state of the system was protected using differential privacy. To improve the system security, an event-triggered set-membership estimation method with privacy-preserving was proposed. Meanwhile, the system stability was analyzed. Finally, it was indicated through simulation experiments that the estimator could be successfully applied, and the introduction of DETS improves the sustainable utilization of resources. This could be followed by research in the incorporation of cyber attacks.

\section*{Acknowledgments}
This work was supported in part by the National Natural
Science Foundation of China under Grant 61903172, Grant 61877065 and Outstanding Youth Innovation Team Project of Shandong Higher Education Institution under Grant 2021KJ042.

\newpage

\end{document}